\documentclass[doublecol]{epl2}

\usepackage{amsmath}
\usepackage{amssymb}
\usepackage{graphicx}
\usepackage{bm}
\usepackage{here}
\usepackage{color}
\usepackage{ulem}
\usepackage{MnSymbol}
\makeatletter
\newcommand{\CustumHeader}[1]{
 \def\epl@stylemark{\hbox to0pt{\hskip0em \vbox to 0pt{\vss \hbox{%
         {#1} %
       }\vskip6ex}\hss}}}
\makeatother

\CustumHeader{}

\def\nn{\nonumber}
\def\bH{\begin{Huge}}
\def\eH{\end{Huge}}
\def\bL{\begin{Large}}
\def\eL{\end{Large}}
\def\bl{\begin{large}}
\def\el{\end{large}}
\def\beq{\begin{eqnarray}}
\def\eeq{\end{eqnarray}}

\def\eps{\epsilon}

\def\e{{\rm e}}

\def\u{\cup}

\def\p{p}
\def\q{q}
\def\P{P}
\def\Q{Q}
\def\kbar{\kappa}
\def\muu{\sqrt{\eps}/\kbar}

\def\Heff{H_{{\rm eff}}}
\def\u{u^{(M)}}

\def\Con{{\rm Con}}
\def\nmax{n_{{\rm max}}}

\def\njat{n^*}
\def\>{\rangle}
\def\<{\langle}

\title{Instanton-noninstanton transition in
nonintegrable tunneling processes: A renormalized
perturbation approach}
\shorttitle{Instanton-noninstanton transition in
nonintegrable tunneling processes}
\author{Akira  Shudo\inst{1}\thanks{E-mail: \email{shudo@tmu.ac.jp}} 
\and Yasutaka Hanada\inst{1}\thanks{E-mail: \email{hanada-yasutaka@ed.tmu.ac.jp}} 
\and Teruaki Okushima\inst{2}\thanks{Email: \email{okushima@isc.chubu.ac.jp}}
\and Kensuke S. Ikeda\inst{3}\thanks{Email: \email{ahoo@ike-dyn.ritsumei.ac.jp}}
}
\shortauthor{A. Shudo \etal}

\institute{
\inst{1}{Department of Physics, Tokyo Metropolitan University,
Minami-Osawa, Hachioji, Tokyo 192-0397, Japan}\\
\inst{2}{
Science and Technology Section, General Education Division, College of Engineering, 
Chubu University, Matsumoto-cho, Kasugai, Aichi 487-8501, Japan}\\
\inst{3}{College of Science and Engineering, Ritsumeikan University
Noji-higashi 1-1-1, Kusatsu 525, Japan}
}


\date{\today}
\abstract{
The instanton-noninstanton (I-NI) transition in the tunneling process, which 
has been numerically observed in classically nonintegrable quantum maps, can be
described by a perturbation theory based on an integrable Hamiltonian
renormalized so as to incorporate the integrable part of the map.
The renormalized perturbation theory is successfully applied to the two quantum maps, 
the H\'enon and standard maps. In spite of different nature of tunneling in 
the two systems, the I-NI transition exhibits very common characteristics. 
In particular, the manifestation of I-NI transition is obviously explained by 
a remarkable quenching of the renormalized transition matrix element. 
The enhancement of tunneling probability after the transition 
can be understood as a sudden change of the tunneling mechanism 
from the instanton to quite a different mechanism 
supported by classical flows just outside of the stable-unstable manifolds of 
the saddle on the top of the potential barrier.
}

\begin{document}
\maketitle

Poincar\'e proved that almost all the classical Hamiltonian systems are nonintegrable. 
The tunneling process in classically integrable systems is described 
almost completely by the instanton theory
and is understood in terms of classical trajectories \cite{Schulmann}.
However, the theory of tunneling in classically nonintegrable systems remains far 
from complete, and is still one of the unsolved fundamental problems in theoretical 
physics. 
It has been more than
two decades since the study of nonintegrable tunneling
starts \cite{LB,BTU,TU,Hensinger,SOR,Ulmmo} (see recent progress in \cite{Creagh,DynamicalTun}).
In the energy domain approach, the application 
of the Herring's formula to nonintegrable systems 
\cite{Wilkinson,Creagh1}, hybrid approaches combining pure quantum 
and semiclassical theory, which was applied to explaining chaos-assisted 
tunneling \cite{BTU,PN,Dresden} and resonance-assisted tunneling \cite{Ulmmo,DresdenSchu}, 
and so on have been proposed.
On the other hand, in the time domain approach, the complex domain semiclassical 
theory has revealed a crucial role of the complexified 
stable-unstable manifold mechanism, which implies the importance
of Julia sets \cite{ShudoIkeda,ShudoIshiiIkeda,TakahasiIkeda}. 
However, the relation and correspondence between theories and numerical 
investigations are not yet clear enough and there is no unified view.

Most of works have been devoted to the so-called quantum maps which model
the essence of nonintegrable feature of quantum dynamics in a simple minded way. 
A very important result which has been reported in several
works is that even a very weak nonintegrable perturbation leads to a remarkable 
enhancement of the tunneling rate 
\cite{Ulmmo,Roncaglia,Fishman,Dresden,DresdenSchu,Mouchet,ShudoIkeda2}. 
This transition is a common feature observed for quantum maps
but its basic origin is still controversial and is not understood
completely. 

Wave functions of non-integrable systems are 
localized on classical tori in the nearly-integrable regime 
according to the well-known Einstein-Brillouin-Keller (EBK) quantization rule. 
A remarkable feature of the invariant torus in nonintegrable 
system is that it breaks at a border called the natural boundary
on which singularities are densely accumulated \cite{GreenPercival}.

It was conjectured that the natural boundary interrupts the instanton, thereby 
may influence the tunneling rate \cite{Creagh}. 
Indeed, clear evidence has been demonstrated recently manifesting 
that the remarkable enhancement of the tunneling rate is closely connected 
with the interruption of instanton due to the natural boundary 
\cite{ShudoIkeda2}. We call the first transition from the instanton tunneling
to some unknown type of tunneling as the instanton-noninstanton (I-NI) 
transition, which occurs in a notable manner in quantum maps.
In the present paper, by introducing a perturbation theory based upon 
maximal renormalization of the integrable part, we show clearly that 
a different tunneling process supported by classical flow 
just outside of the stable-unstable manifolds of the saddle
at the top of the potential barrier induces the transition to the NI regime.

In what follows, we take the symmetric form of the quantum map
$U  =  \e^{-i\P^2/4\hbar}\e^{-i\eps V(\Q)/\hbar}\e^{-i\P^2/4\hbar}$ 
(or $U=\e^{-i\eps V(Q)/2\hbar}\e^{-i P^2/2\hbar}\e^{-i\eps V(Q)/2\hbar}$). 
Redefining the effective Planck constant as $\kbar=\hbar/\sqrt{\eps}$ 
and the new set of conjugate operators $\p=-i\kbar d/d\q$ and $\q=\Q$, 
then $U$ becomes a convenient form in our approach 
\begin{eqnarray}
\label{U}
U  =  \e^{-i\sqrt{\eps}p^2/4\kappa}\e^{-i\sqrt{\eps}V(\q)/\kappa}\e^{-i\sqrt{\eps}\p^2/{4\kappa}}.
\end{eqnarray}%

\begin{figure}[htbp]
\begin{center}
\includegraphics[width=0.45\textwidth]
{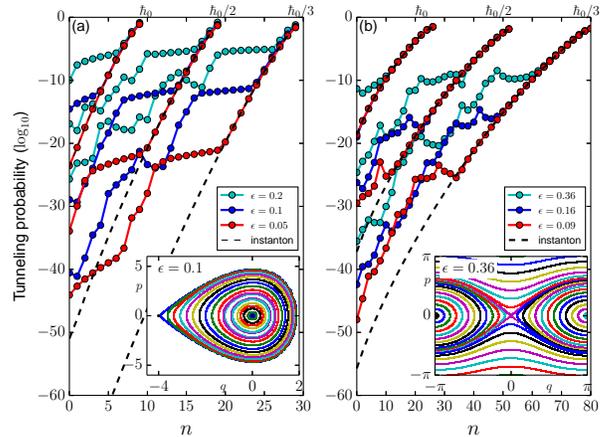}
\caption{\label{FigTunprbHS} 
The tunneling probability (in $\log_{10}$) for the eigenstates of eq. (1)
vs the quantum number $n$ (less than $\nmax$) for  
(a) the H\'enon and (b) the standard map with various values of $\hbar$ 
in the unit of $\hbar_0$ and the nonlinear parameter $\eps$. 
The tunneling probability is computed in the asymptotic
region ($\Q\ll -4$) for the H\'enon map and at the 
the potential top ($Q=0$) for the standard map
(only the even parity eigenfunctions in the 
$q$-representation are considered).
Here $\hbar_0=0.63\sqrt{\eps}$ for the H\'enon map and 
$\hbar_0=\frac{3\pi}{25}\sqrt{\eps}$
for the standard map. Insets shows the phase space portrait 
of classical maps with $\eps=0.1$ and $\eps=0.36$ for the H\'enon
and standard maps, respectively.
}
\end{center}
\end{figure}

Figs. \ref{FigTunprbHS}(a) and (b) are typical examples of tunneling characteristics
computed for two typical quantum maps, namely the H\'enon and standard maps. 
The figures show the representative tunneling probability as a function of the 
quantum number, and they exhibit a typical feature of the I-NI transition.
Here the quantum maps have the potential functions $V(\Q)=2\Q^2+\Q^3/3$ 
(the H\'enon map) and $\cos Q$ (the standard map), respectively. 
Although these examples are nonintegrable, we here examine weakly nonintegrable 
regimes with small $\eps$ as displayed in the insets of fig. \ref{FigTunprbHS}. 
Therefore, the classical invariant tori still remain 
according to the celebrated Kolmogorov-Anold-Moser (KAM) theory, 
and good quantum numbers can be assigned to each quantum eigenstate 
following the EBK quantization rule.
%

In the limit of vanishing nonintegrability, both systems are approximated
by one-dimensional systems with potential barriers which classically bound
the eigenstates but quantum mechanically allow to tunnel by the instanton
mechanism as is usual in one-dimensional systems, which will be discussed later.  
However, as the quantum number $n$ decreases from the highest exited 
state with the energy just below the potential barrier height, 
a transition occurs at a characteristic quantum number,         
denoted hereafter by $n_{\rm c}$, and the tunneling probability suddenly deviates 
from the instanton probability forming a {\it plateau}. 
The common feature seems to be quite paradoxical in the 
sense that the low lying states distant from the saddle point 
located on the barrier top, which is the very origin of the classical nonintegrability 
exhibit non-instanton tunneling, while the highly excited 
states close to the saddle point obey the instanton tunneling. Moreover,
in both examples the number of the eigenstates in the
instanton regime $\nmax-n_{\rm c}$ is insensitive to $\hbar$, where $\nmax$
is the maximum quantum number of the classically bounded states
inside the potential well.

To understand the I-NI transition the most natural way is 
to apply a perturbation theory based on an integrable limit 
which has the instanton as the tunneling mechanism.
The crudest integrable approximation of symmetrized $U$ in
the small limit of the nonlinear parameter $\eps$ is
$U_1 = \e^{-i\muu H_1}$ 
where $H_1=\p^2/2+V(\q)$, but the difference 
$|U-U_1|\sim \eps^2/\kbar$ is too large to afford any significant 
result for the exponentially small tunneling effect. We therefore  
develop a systematic expansion with respect to the smallness parameter $\eps$
which renormalizes the integrable part of $U$ into a single 
effective Hamiltonian as much as possible. 
A possible candidate to achieve this 
is the Baker-Hausdorff-Campbell (BHC)
expansion, 
which approximates the product of exponential operators (\ref{U}) 
systematically in terms of a single exponential operator expressed by
an effective Hamiltonian, which coincides with $H_1$ in the lowest-order approximation.

Below, we apply a renormalized perturbation theory to the H\'enon 
and standard maps. They are both simplest classes of maps exhibiting quite different 
nature of tunneling, which is understood by the simplest one-dimensional 
Hamiltonian $H_1=\p^2/2+V(\q)$:
the H\'enon map has the cubic potential with the bottom at $\q=0$ and tunneling
is an irreversible transport of probability toward $\q=-\infty$ going over the
barrier at $\q=-4$.
For the standard map the $4\pi$ periodic boundary condition is imposed 
in the $q$ direction. Then its cosine potential gives two symmetric 
valleys with bottoms at $\q=\pm\pi$ and separated by two symmetric 
barriers at $\q=0$ and $\q=2\pi$. 

We can show that the BHC expansion of $U$ leads to an 
effective polynomial Hamiltonian 
\begin{eqnarray}
\label{U2}
&&\Heff^{(M)}=\sum_{\ell=1,3,..,M}(-\eps)^{(\ell-1)/2}H_{\ell}\\
\nonumber
&&{\rm with}~~H_\ell = \sum_{k,i,j=0}^{k_\ell,i_{\ell k},j_{\ell k}} \kbar^{k} C(\ell,k,i,j) \p^{i}\q^{j}, 
\end{eqnarray}
where $C(n,k,i,j)$ is the set of 
coefficients of $O(1)$, and the terms of $k=0$ provide the classical Hamiltonian. 
Note that only the odd powers of $\eps^{1/2}$ appear in the sum because of 
the symmetrized form of $U$. 
The unitary operator thus induced is given as $U_{M} = \e^{-i\muu \Heff^{(M)}}$. 
We do not introduce any artificial absorbers and/or absorbing boundary conditions,
which may crucially influence the original dynamics \cite{Dresden,DresdenSchu}.

Once the renormalized Hamiltonian is obtained, we straightforwardly develop a perturbation 
theory taking the difference $\Delta U_M = U-U_M=O(\eps^{(M+1)}\muu)$
as the perturbation. Then the lowest-order perturbative eigenfunction is given as 
\begin{eqnarray}
\label{pert} 
\nonumber 
|\Psi^{(M)}_n\rangle &=&|\u_n\rangle+|\Delta \Psi^{(M)}_n\rangle, \\
|\Delta \Psi^{(M)}_n\rangle &=&
\sum_{j}\frac{\langle\u_j|\Delta U_M|\u_n\rangle}{\e^{-i\muu E^{(M)}_n}-\e^{-i\muu E^{(M)}_j}}|\u_j\rangle, 
\end{eqnarray}    
where $|\u_n\rangle$ and $E^{(M)}_n$ are respectively the eigenfunction and the energy eigenvalue 
of the 1D integrable Hamiltonian $\Heff^{(M)}$.

If $\eps$ is not so large,  there appear two sorts of fixed points in classical phase space:
stable and unstable fixed point $(0,0)$ and $(-4,0)$ for the H\'enon map,
and $(\pm \pi, 0)$ and $(0,0)$ for the standard map respectively.
As shown in the insets of fig. \ref{FigTunprbHS}, 
KAM tori predominate phase space in both cases and KAM regions are 
encircled by the stable manifold $W^s$ and
the unstable manifold $W^u$ of the saddle fixed point.
KAM tori support quantum eigenstates, each of which 
satisfies the EBK quantization condition 
$\int \p d\q/2\pi=(n+1/2)\kbar~~(n=0,1, \cdots, \nmax)$. 
They all have finite tunneling life-times (H\'enon map) 
or tunneling splittings 
(standard map) due to the tunneling via the instanton trajectory, 
which is very well approximated by the lowest order instanton solution 
$\p=\sqrt{2(V(\q)-E_n)}$ or $H_1({\rm i}\p,\q)=E_n$ ($E_n$ is the quantized energy).
A notable feature of the renormalized Hamiltonian $\Heff^{(M)}(\p,\q)$ 
is that tunneling tails of eigenfunctions $\<\q|\u_n\>$ do not change very significantly 
even if one increases the order of renormalization $M$. 
However, according to eq.(\ref{pert}), 
the application of perturbation changes drastically the tunneling tail of eigenstates 
with increase in $M$.

\begin{figure}[htbp]
\begin{center}
\includegraphics[width=0.45\textwidth]
{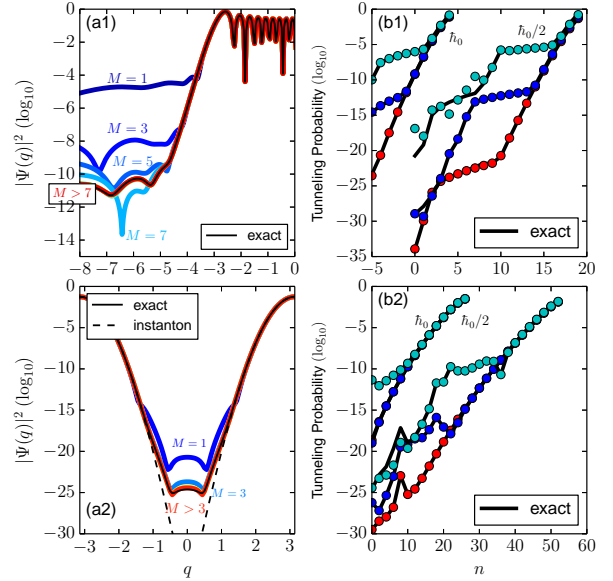}
\caption{\label{FigTunprbExactPert} 
The convergence of tunneling tail to the exact numerical one (black) 
with increase in $M$ 
in case of (a1) $n=13$-th excited state of the H\'enon map 
with $\eps=0.1$ and $\hbar=\hbar_0/2$, and 
(a2) the ground state of standard map with $\eps=0.36$ and $\hbar=\hbar_0/2$. 
The exact and higher order renormalized perturbative results are not 
distinguishable with each other. 
Right-hand panels show the probability amplitude of tunneling tail of 
exact numerical diagonalization (black lines) 
and of perturbation calculation as a function of quantum number (less than $\nmax$) 
in case of (b1) the H\'enon map with the order $M=21$, and 
(b2) the standard map with the order $M=7$.  
In (b2) the quantum number is shifted by -5 for $\hbar_0$, 
and so $n=-5$ means $n=0$ for $\hbar_0$.
}
\end{center}
\end{figure}

In fig. \ref{FigTunprbExactPert}(b) we show a typical result of renormalized 
perturbation theory. For lower-order approximation, $U-U_{M}$ 
is too large to control correctly the
exponentially small tunneling component and so the perturbative solution yields
quite absurd results. But as the order $M$ of renormalization increases 
the perturbative solution converges to the exact 
eigenfunction $|\Psi_n\>$ obtained by numerical diagonalization
and reproduces even the complicated oscillations at the tunneling tail
as demonstrated in figs. \ref{FigTunprbExactPert} (a1) and (a2). 
The higher-order renormalized perturbation calculation also succeeded in 
reproducing the exact tunneling probability in a rather 
wide regime including both I (instanton) and I-NI transition regions.
(See figs. \ref{FigTunprbExactPert} (b1) and (b2)).

The success of the perturbation theory means that the origin of the 
I-NI transition may be resolved into integrable bases:
according to eq. (\ref{pert}) 
we define the {\it contribution spectrum} representing the amount of 
contribution from the $j$-th eigenstate of the integrable model to the 
$n$-th perturbative eigenfunction: 
\begin{eqnarray}
\lefteqn{\Con^{(M)}_{j\to n} \equiv 
\llangle|\< \q |\u_j \> |\rrangle |\< \u_j|\Psi^{(M)}_n \> | 
\simeq} \nn\\ 
&\begin{cases} 
\llangle |\< \q|\u_j \> |\rrangle |\<u_j|\Delta \Psi^{(M)}_n\>|
& (\text{if } j\neq n) \\[1mm] 
\llangle|\< \q|\u_n \> |\rrangle  & (\text{if }j=n)  
\end{cases}
\end{eqnarray}
where $\llangle \dots \rrangle$ means to take average over a range of $q$ in the asymptotic region
(H\'enon map) or around $q=0$ (standard map).

\begin{figure}[htbp]
\begin{center}
\includegraphics[width=0.45\textwidth]
{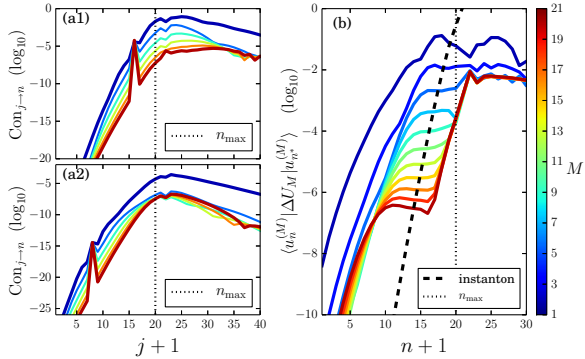}
\caption{\label{FigTunprbConPlateau} 
Typical contribution spectra $\Con^{(M)}_{j\to n}$ of (a1) before 
and (a2) after the I-NI transition. (a1) and (a2) respectively 
correspond to $n=15$ and $n=7$ of the H\'enon map with $\eps=0.1$ and $\hbar=\hbar_0/2$
(see fig. \ref{FigTunprbExactPert}(b1)).
The renormalization order grows as $M=1,5,9,13,17,21$. 
(b) The major transition matrix element 
$|\langle\u_n|\Delta U_M|\u_{\njat}\rangle|$ vs $n$ 
for the same H\'enon map, where $\njat=21=\nmax+2$.
$M$ is increased from 1 to 21.
The black doted curve denotes the instanton peak amplitude 
$\llangle|\langle q|\u_n\rangle|\rrangle$ of the $n$-th state.
The vertical axis is drawn in the $\log_{10}$ scale. 
}
\end{center}
\end{figure}

We discuss closely how the renormalized perturbation theory describes the I-NI transition
by taking the contribution spectrum of the H\'enon map as an example.
Figure \ref{FigTunprbConPlateau}(a) shows how the contribution spectrum 
$\Con^{(M)}_{j\to n}~~(0\leq j \leq \nmax)$ varies with 
the order of renormalization
$M$ at two representative quantum numbers $n$ before and after the I-NI transition. 
A quite interesting fact is that for $M=1$ the largest contribution comes 
from a broad peak centered at $j=\njat$, where $\njat$ is 
the number of a quantum state just 
above the threshold of dissociation $n=\nmax$. However, with increase in the renormalization
order $M$, the transition matrix elements 
$|\< \u_j|\Delta U_{M}|\u_n\>|$
are remarkably quenched, and the peak at $j=\njat$ is reduced in the logarithmic scale
and finally overwhelmed by the very sharp peak at $j=n$, which means that the contribution 
from instanton of the integrable basis $\<q|\u_n\>$ is predominant in the tunneling process. 
On the other hand, at $n$ less than a critical quantum number $n_{\rm c}$ 
the quenching of the transition matrix element can no longer reduce 
the peak at $j=\njat$ less than the instanton peak, and the contribution
to tunneling is attributed to the broad peak around $j=\njat$. The competition between
the two peaks explains the characteristics of I-NI transition.
It should be emphasized that without remarkable quenching of 
the transition matrix element by renormalization the instanton phase is absent 
and the I-NI transition cannot be observed. 

%
%

Since the above threshold state $|\u_{\njat}\rangle$ is connected with $q=-\infty$ 
by real classical paths, $|\<q|\u_{\njat}\>|\sim O(1)$ and so the peak strength $\Con^{(M)}_{\njat\to n}$ 
at $\njat$ is approximated by the transition matrix element $|\<\u_n|\Delta U_M|\u_{\njat}\>|$.
Figure \ref{FigTunprbConPlateau} (b) plots $|\<\u_n|\Delta U_M|\u_{\njat}\>|$ at just above 
the threshold for various values of $M$. The instanton peak strength 
$\llangle| \< q| \u_{n}\rangle|\rrangle$ is also shown as a function
of $n$, which does not significantly depend on $M$ as mentioned above. 

With increase in $M$, the curve $|\langle\u_n|\Delta U_M|\u_{\njat}\rangle|$ vs $n$ 
is largely deformed to show a very characteristic structure: it 
decreases steeply as $n$ decreases below $\nmax$, but it reaches 
a definite plateau and then it decreases again steeply. It is just 
on the plateau that the curve $|\<\u_n|\Delta U_M|\u_{\njat}\>|$ vs $n$
intersects with the instanton peak curve,  which means that tunneling 
using the transition to the state $n=\njat$ overwhelms the instanton 
tunneling and so the intersection determines  $n_{\rm c}$. As $M$ increases, the height of 
plateau decreases rapidly and reaches finally to a limit, which
causes the instanton region $n_{\rm c}<n<\nmax$ to grow from a null region 
to a finite region with the width proportional to $\sqrt{\eps}$. 
Below $n_{\rm c}$, the transition matrix element controls tunneling and so the 
tunneling probability follows the plateau structure, which explains the 
characteristic plateau of the tunneling amplitude seen in 
fig.\ref{FigTunprbExactPert}.

The origin of the transition to a plateau-like characteristics of the tunneling rate 
from the instanton tunneling rate reported in preceding works
can therefore be attributed to the formation of the plateau of the transition matrix element
and a drastic decrease of plateau height in higher-order renormalization. 
The I-NI transition in the standard map in fig.\ref{FigTunprbHS}(b) follows the same scenario
\cite{FuturePublications}.


We note that if one shifts each curve in Fig. \ref{FigTunprbHS} 
horizontally such that the maximal quantum 
number $n_{\rm max}$ for each curve coincides with each other, 
they all show very similar characteristics, and are insensitive to the effective 
Planck constant $\kappa$. 
Such a feature can hardly be explained by classical 
objects {\it e.g.,} nonlinear classical resonances, which will be discussed in detail in 
our forthcoming papers \cite{FuturePublications}.

\begin{figure}[htbp]   
\begin{center}%
\includegraphics[width=8cm]
{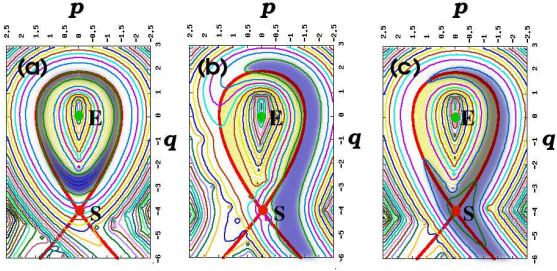}
\caption{\label{FigHusimiWsWu}
Husimi plots of the PTC for the eigenfunctions of
the H\'enon map with $\eps=0.10$ and $\hbar=\hbar_0/2$ (see fig. \ref{FigTunprbExactPert}(b1)).
(a) before ($n=15$), and (b) after ($n=13$) the I-NI transition, and (c) is the 
eigenstate at the edge of plateau ($n=7$).
(a) and (c) correspond to fig. \ref{FigTunprbConPlateau}(a1) and (a2), respectively.
Contours are plotted in log scale, and the shaded regions indicates the highest levels.
The red curve is the $W^s$-$W^u$ complex of the saddle $S$ (red circle), and
the green dot $E$ denotes the stable fixed point.
The Husimi levels of (a) coincide entirely with the eigenfunction of $|\u_{15}\>$ of $\Heff^{(M)}$, 
which has its major body inside of the $W^s$-$W^u$ complex, 
while in (b) and (c) the major body winds around  $W^s$, passes nearly to
$S$ and is blown to $-\infty$ following the $W^u$. 
The maximum level of contour is $10^{0.5}$ in (a), 
suddenly drops to $10^{-4.5}$ in (b) after the I-NI transition }
\end{center}
\end{figure}

Finally, we elucidate the classical dynamical significance of the plateau region
which is characteristic after the I-NI transition.
The above analyses tell us that the states composed of the self-component
$j=n$ and the ones forming the broad peak around $j=\njat$ play as the 
principal component. We define here the principal component contributing to 
tunneling (PCT) and observe the phase space Husimi-plot to investigate 
the classical interpretation for the I-NI transition. The PCT is defined as the 
projection of $|\Psi^{(M)}_n\rangle$ onto the principally contributing subspace 
constructed as follows. Let $|\u_{j}\rangle$ be eigenfunctions rearranged 
in descending order of the magnitude $\Con_{j \to n}$ in the contribution spectrum, 
and consider the $K-$dimensional subspace spanned by $|\u_{j}\rangle ~~ (1\le j \le K)$. 
Let $K_{\rm min}$ be the minimum number of $K$ which makes the relative distance between 
the vector $|\Psi^{(M)}_n\>$ and its projection onto the above introduced subspace less
than a small enough threshold value $r_{th}~(\ll 1)$, namely, the minimal $K=K_{\rm min}$ 
such that
$$
\Big\llangle\frac{|\<q|\Psi^{(M)}_n\>-\sum_{j=0}^{K}\<q|\u_{j}\>\<\u_{j}|\Psi^{(M)}_n\>|}{|\<q|\Psi^{(M)}_n\>|} \Big\rrangle <r_{th}
$$
Then the space spanned by $|\u_{j}\rangle ~~ (1\le j < K_{\rm min})$
constitutes the principally contributing subspace to tunneling. 
We take $r_{th}=0.2$, for example.
We have to emphasize that $M$ must be taken sufficiently 
large ($M \geq 13$ in practice) in order that the PCT is significant.

The Husimi plots of the PCT before and after the I-NI transition are 
depicted in fig. \ref{FigHusimiWsWu}. The shadowed region indicates 
the region with the highest probability level. 
In the instanton regime the PCT coincides almost with the eigenfunction 
$|\u_n\>$ of the unperturbed integrable Hamiltonian, and the Husimi plot 
of PCT in fig. \ref{FigHusimiWsWu}(a) indeed traces the classical invariant 
circle shown as the shadowed region, and the tunneling component is 
represented by the monotonously decaying contours encircling the shaded quantized 
invariant circle. In particular, when observed in the $q$-coordinate, the tunneling 
tail is the line $p=0$ passing across the saddle $S$, which is nothing more than 
the instanton. 

On the other hand, fig. \ref{FigHusimiWsWu}(b) indicates that a drastic change occurs 
in PCT when the transition to NI happens: the main component, characterized 
by almost the same probability levels, runs along the stable manifold $W^s$, 
attracted and repelled by the saddle $S$, finally runs away toward $q=-\infty$ along 
the unstable manifold $W^u$, which manifests that the PCT contributing to the 
tunneling tail represents classical flows just outside of the $W^s$-$W^u$ complex. 
%
%
The PCTs for all the eigenstates forming the plateau take almost similar 
patterns but they approach more closely to the $W^s$-$W^u$ complex,
as the quantum number shifts to the edge of the plateau. 
Figure \ref{FigHusimiWsWu}(c) shows that just at the eigenstate on the edge 
of the plateau its PCT coincides with the $W^s$-$W^u$ complex. 
As is seen in fig. \ref{FigTunprbHS}, the tunneling amplitude decreases rapidly
when $n$ is less than the plateau edge quantum number.
All the above features are common in the H\'enon and  standard maps, 
which are more completely discussed in \cite{FuturePublications}.

In the present paper we have developed a perturbation theory for nearly 
integrable quantum maps. 
This is based upon an integrable Hamiltonian which is 
constructed by maximally renormalizing the integrable part of the map. 
This was successfully applied to investigate the I-NI transition commonly
observed in nearly integrable quantum maps such as the H\'enon and standard maps. 
A remarkable quenching of the highly renormalized transition matrix elements 
explains the origin of the abrupt change of tunneling characteristics 
at the I-NI transition.  The PCT analysis reveals that 
the tunneling mechanism in the plateau region after the I-NI transition is due to 
the classical flow outside of the $W^s$-$W^u$ complex is 
responsible for the tunneling process after the transition 
and the flow coincides with the $W^s$-$W^u$ complex at the edge of the plateau.
This suggests the crucial role of classical dynamics
related to the $W^s$-$W^u$ complex, and 
elucidating the relation to the complex stable-unstable 
manifolds mechanism based on the complex-domain 
semiclassics is strongly desired.


\acknowledgments
Discussions with Y. Shimizu, K. Takahashi, 
A. B\"acker, R. Ketzmerick, N. Mertig and A. Mouchet are appreciated.
This work was supported by Kakenhi 24340094 based on the tax of Japanese people,
and the authors would like to acknowledge them.  
This was also supported by 
Chubu University Grant (26IIS06AII). 
The authors are very grateful to Shoji Tsuji and Kankikai for using 
their facilities at Kawaraya during this study.



\begin{thebibliography}{0}
\bibitem{Schulmann}
	\Name{Schulman L. S.}
	\Book{Techniques and Applications of Path Integration}
	\Publ{Wiley-interscience}
	\Year{1996}.	
	



\bibitem{Ulmmo} 
	\Name{Brodier O., Schlagheck P. \and Ullmo D.}
	\REVIEW{Phys. Rev. Lett.} {87}{2001}{064101};
	\REVIEW{Ann. Phys.} {300}{2002}{88}.
     The effect of classical nonlinear resonances on tunneling was discussed in
	\Name{Ozorio de Almeida A. M.}
	\REVIEW{J. Phys. Chem.} {88}{1984}{6139}.

	
\bibitem{LB} 
	\Name{Lin W. A. \and Ballentine L. E.}
	\REVIEW{Phys. Rev. Lett.} {65}{1990}{2927}.
	
\bibitem{BTU} 
	\Name{Bohigas O., Tomsovic S. \and Ullmo D.}
	\REVIEW{Phys. Rep.} {223}{1993}{45}. 
	
\bibitem{TU} 
	\Name{Tomsovic S. \and Ullmo D.}
	\REVIEW{Phys. Rev. E} {50}{1994}{145}. 
	

	
\bibitem{Hensinger} 
	\Name{Hensinger W. K. {\it et al.}}
	\REVIEW{Nature} {412}{2001}{52}. 

\bibitem{SOR}
	\Name{Steck D. A., Oskay W. H. \and  Raizen M. G.}
	\REVIEW{Science} {293}{2001}{274}.	

	
	
\bibitem{Creagh}
	\Name{Creagh S. C.} 
	\Book{Tunneling in Complex Systems}
	\Editor{S.Tomsovic} 
	\Publ{Singapore, World Scientific}
	\Year{1998}
	\Page{35}


\bibitem{DynamicalTun}
	\Book{Dynamical Tunneling: Theory and Experiment}
	\Editor{Keshavamurthy S. \and Schlagheck P.}
	\Publ{CRC Press}
	\Year{2011}.

\bibitem{Wilkinson} 
	\Name{Wilkinson M.}
	\REVIEW{Physica D} {21}{1986}{341}.

	
\bibitem{Creagh1}
	\Name{Creagh S. C. \and Finn M. D.}
	\REVIEW{J. Phys. A} {34}{2001}{3792}.


\bibitem{PN} 
	\Name{Podolskiy V. A. \and Narimanov E. E.}
	\REVIEW{Phys. Rev. Lett.} {91}{2003}{263601}.
	
\bibitem{Dresden}
	\Name{B\"acker A., Ketzmerick R.,  L\"ock S., \and Schilling L.}
	\REVIEW{Phys. Rev. Lett.} {100}{2008}{104101};
	\Name{B\"acker A., Ketzmerick R., \and  L\"ock S.}
	\REVIEW{Phys. Rev. E} {82}{2010}{056208}.
	
\bibitem{DresdenSchu} 
	\Name{L\"ock S., B\"acker A., Ketzmerick R. \and Schlagheck P.}
	\REVIEW{Phys. Rev. Lett.} {104}{2010}{114101}.
	
\bibitem{ShudoIkeda} 
	\Name{Shudo A. \and Ikeda K. S.} 
	\REVIEW{Phys. Rev. Lett} {74}{1995}{682}; 
	\REVIEW{Physica D} {115}{1998}{234}.
	
\bibitem{ShudoIshiiIkeda} 
	\Name{Shudo A., Ishii Y. \and Ikeda K. S.}
	\REVIEW{J. Phys. A} {42}{2009}{265101};
	\SAME{42}{2009}{265102}.
	
\bibitem{TakahasiIkeda} 
	\Name{Takahashi K. \and Ikeda K. S.}
	\REVIEW{J. Phys. A} {43}{2010}{192001}.

\bibitem{Roncaglia} 
	\Name{Roncaglia, R., Bonci, L., Izrailev, F.M., West, B. J. \and Grigolini, P.}
	\REVIEW{Phys. Rev. Lett.} {73}{1994}{802}.
	
	
\bibitem{Fishman}
	\Name{Sheinman M., Fishman S., Guarneri I. \and Rebuzzini L.}
	\REVIEW{Phys. Rev A} {73}{2006}{052110}.
	
\bibitem{Mouchet} 
	\Name{Mouchet A.}
	\REVIEW{J. Phys. A} {40}{2007}{F663}.

\bibitem{ShudoIkeda2}
	\Name{Shudo A. \and Ikeda K. S.}
	\REVIEW{Phys. Rev. Lett.} {109}{2012}{154102}. 

\bibitem{GreenPercival}
    \Name{Greene M. \and Percival I. C.}
    \REVIEW{Physica}{3D}{1981}{530}.

\bibitem{FuturePublications}
      \Name{Hanada Y., Shudo A. \and  Ikeda K. S.}
      in preparation;
      \Name{Shudo A., Hanada Y., Okushima T. \and  Ikeda K. S.}
      in preparation.

\end{thebibliography}
\end{document}